\documentstyle[12pt,epsfig]{article}

\begin{document}
\begin{flushright}
hep-ph/0508012 \\
RAL-TR-2005-013 \\
6 Sep 2005 \\
\end{flushright}
\vspace{2 mm}
\begin{center}
{\Large 
Covariant Extremisation of Flavour-Symmetric} 
\end{center}
\vspace{-8mm}
\begin{center}
{\Large Jarlskog Invariants and the Neutrino Mixing Matrix}
\end{center}
\vspace{3mm}
\begin{center}
{P. F. Harrison\\
Department of Physics, University of Warwick,\\
Coventry, CV4 7AL. UK \footnotemark[1]}
\end{center}
\begin{center}
{and}
\end{center}
\begin{center}
{W. G. Scott\\
CCLRC Rutherford Appleton Laboratory,\\
Chilton, Didcot, Oxon OX11 0QX. UK \footnotemark[2]}
\end{center}
\vspace{3mm}
\begin{abstract}
\baselineskip 0.6cm
\noindent
We examine the possibility that
the form of the lepton mixing matrix
can be determined by extremising 
the Jarlskog flavour invariants
associated, eg.\ with 
the commutator ($C$) of the lepton mass matrices.
Introducing a strictly covariant approach,
keeping masses fixed and
extremising the determinant (Tr $C^3/3$) leads to maximal $CP$ violation,
while extremising the sum of the $2 \times 2$ principal minors ($-{\rm Tr} \, C^2/2$),
leads to a non-trivial mixing with zero $CP$ violation.
Extremising, by way of example, a general linear combination 
of two $CP$-symmetric invariants together, 
we show that our procedures can lead 
to acceptable mixings and to non-trivial predictions, 
eg.\ $|U_{e3}| \simeq \sqrt{2}/3 \;
\sqrt{\Delta m_{12}^2/\Delta m_{23}^2}(1-m_{\mu}/m_{\tau})^2 \simeq 0.07$.

\end{abstract}
\begin{center}
{\em To be published in Physics Letters B}
\end{center}
\footnotetext[1]{E-mail:p.f.harrison@warwick.ac.uk}
\footnotetext[2]{E-mail:w.g.scott@rl.ac.uk}
\newpage
\baselineskip 0.6cm

\noindent {\bf 1. Introduction} \\

\noindent 
It is now almost twenty years since Jarlskog \cite{jinv}
first alerted us to the inherent U(3) invariance representing the freedom
in any weak basis \cite{wbas}
to transform fermion mass matrices in the generation space,
whilst keeping the charged-current weak-interaction diagonal and universal,
ie.\ whilst always staying in a weak basis.
While all observables 
(eg.\ masses and mixing angles)
must of course be Jarlskog-invariant,
we focus here on flavour-symmetric invariants,
ie.\ those invariants without flavour indices of any kind
(ie.\ involving no flavour projection operators in their definition).

While it has long been clear \cite{binv}
that any fundamental laws underlying the masses and mixings
should be Jarlskog-covariant (as defined below),
this principle may not 
always be easy to respect in practice.
In this paper we will show that
extremisation of flavour-symmetric Jarlskog invariants \cite{hs1} \cite{rod1}
leads naturally to Jarlskog-covariant constraints,
which (as we shall see) can even have viable phenomenology.

The archetypal example of a flavour-symmetric Jarlskog invariant
is the famous determinant of the commutator \cite{jcp}, 
which for the charged-lepton ($L$) and neutrino ($N$)
mass matrices (taken here to be hermitian) may be written:
\begin{eqnarray}
\hspace{0.4cm}
{\rm Det } \; C \, = \, {\rm Tr} \; C^3/3 \, = \;
\, {\rm Tr} \; i[L,N]^3/3 \, = \;
-2 \; \; {\rm Det} \; {\rm diag} (\Delta_l) \; \; {\rm Det} \; {\rm diag} (\Delta_{\nu}) 
\; J \hspace{0.5cm} \label{detc1} \\
\Delta_l^T=(m_{\mu}-m_{\tau},\, m_{\tau}-m_{e}, \, m_{e}-m_{\mu}), 
\hspace{0.5cm}
\Delta_{\nu}^T=(m_{2}-m_{3},\, m_{3}-m_{1}, \, m_{1}-m_{2}) \label{detc2}
\end{eqnarray} 
where $C := -i[L,N]$, with $m_e$, $m_{\mu}$, $m_{\tau}$ the charged-lepton masses
and $m_1$, $m_2$, $m_3$ the neutrino masses, 
and with the invariant measure of $CP$ violation \cite{jcp} given by:
\begin{equation}
J=c_{12} \,  s_{12} \; c_{23} \, s_{23} \; c_{13}^2 \, s_{13} \; s_{\delta}
=(1-s_{12}^2)^{\frac{1}{2}} \, s_{12} \; (1-s_{23}^2)^{\frac{1}{2}} \, s_{23}
(1-s_{13}^2) \, s_{13} \; s_{\delta}. \label{jpdg}
\end{equation}
Extremising Eq.~\ref{jpdg},
with respect to the standard PDG \cite{pdg} variables 
$s_{12}$, $s_{23}$, $s_{13}$ and $\delta$  
is known \cite{sqt3} to lead to 
$s_{12}=1/\sqrt{2}$, $s_{23}=1/\sqrt{2}$, 
$s_{13}=1/\sqrt{3}$ and $\delta=\pi/2$,
corresponding to maximal $CP$ violation,
ie.\ to trimaximal mixing \cite{hps1}. 

It should be remarked that
the assumption 
that the mass matrices are hermitian (see above)
may in fact be realised in two distinct ways: 
one may simply imagine applying a suitable transformation 
to the right-handed fields 
(which are anyway inert to the charged-current weak-interaction)
bringing $L$ and $N$ into hermitian form,
or one may instead everywhere re-interpret $L$ and $N$ as hermitian squares 
of masss matrices $L \rightarrow LL^{\dagger}$, $N \rightarrow NN^{\dagger}$,
operating between left-handed fields.
In the latter case, masses need to be replaced by masses-squared throughout 
($m_{l/\nu} \rightarrow m_{l/\nu}^2$, eg.\ in Eqs.~\ref{detc1}-\ref{detc2}).
In this paper, in order to circumvent this as yet unresolved ambiguity, 
we shall give results, where appropriate, 
for both the linear and quadratic cases. 

Of course we now know that lepton mixing is not trimaximal,
it being actually much closer to the so-called tri-bimaximal 
form \cite{tbm1} \cite{tbm2}.
The question then arises, 
as to whether extremising some other invariant
(or combination of invariants) aside from ${\rm Tr} \, C^3$,
might perhaps yield a more realistic mixing prediction.

In this paper we seek to
kindle interest in these kinds of questions, 
and we start out (Section~2) by re-deriving 
the results for ${\rm Tr} \, C^3$ above,
introducing and bringing to bear 
a strictly covariant (basis-independent) approach.
We focus here on the leptonic case:
in the context of extremisation, 
quasi-maximal mixing in the lepton sector 
looks a priori 
more comprehensible than 
anything involving the quarks
(which are anyway subject to larger radiative corrections,
eg.\ dependent on the top mass).

We proceed (Section~3)
with a similar covariant extremisation of
the only other independent invariant depending on $C$ alone,
ie.\ ${\rm Tr} \, C^2$.
The mixing prediction 
is now $CP$-conserving and,
whilst not phenomenologically viable as it stands,
is non-trivial,
with even a suggestive resemblance
to some previously proposed lepton-mixing ansatze.

Finally, taking a particular linear combination of 
two $CP$-symmetric invariants as an illustrative example,
we show (Section~4) how covariant extremisation 
of slightly more general combinations of invariants
can readily lead to realisic mixing predictions.

Our ``Perspective'' section (Section~5) 
simply consolidates what we have learned,
also briefly pointing forward to what one 
might hope to learn in future studies,
eg.\ in terms of incorporating $CP$ violation,
dropping mass constraints etc.\\

\noindent {\bf 2. Extremising the Jarlskog Determinant, Tr}
\begin{boldmath}$C^3/3$\end{boldmath} \\

\noindent 
One might argue that the PDG variables are arbitrary,
eg.\ it matters (Eq.~\ref{jpdg}) that we chose to extremise with respect to $\delta$ 
and not $s_{\delta}=\sin \delta$. 
We choose to
extremise with respect to the Yukawa couplings themselves
(ie.\ with respect to the mass matrices) 
using some established results \cite{horn} 
on differentiation of matrix traces with respect to matrix variables
(see Appendix A).
In terms of matrix derivatives $\partial_L := \partial/\partial L$ etc.,
the extremisation conditions for, eg.\ ${\rm Tr } \, C^3$ 
may then be written (see Appendix A)
in a manifestly Jarlskog-covariant (ie.\ basis-independent) way:
\begin{eqnarray}
\partial_L \, {\rm Tr} \; C^3/3 &=& -i[N,C^2]^T = 0 \label{exsl} \\
\partial_N \, {\rm Tr} \; C^3/3 &=& +i[L,C^2]^T = 0 \label{exsn},
\end{eqnarray} 
in that both sides of the equation are evidently
form-invariant under an arbitrary $U(3)$ transformation
$L\rightarrow U(3)L \, U(3)^{\dag}$, 
$N\rightarrow U(3)N \, U(3)^{\dag}$,
ie.\ we would 
obtain the same equations performing the 
extremisation in any basis. 
When the masses are to be held fixed,
the zeros on the RHS of Eqs.~\ref{exsl}-\ref{exsn} 
must be replaced by 
arbitrary matrix polynomials in $L^T$ and $N^T$ respectively, 
representing the dependence on unknown Lagrange multipliers (see below),
with the manifest covariance clearly maintained.

The covariance property allows us to solve our equations in any basis, 
so we choose a basis convenient to us 
where the charged-lepton mass matrix is diagonal
and where all imaginary parts 
of the off-diagonal elements of the neutrino mass matrix  
are equal (the `epsilon' basis \cite{simp}).
The neutrino mass matrix may then be written:
\begin{eqnarray}
     \matrix{  \hspace{0.0cm} e \hspace{0.6cm}
               & \hspace{0.4cm} \mu \hspace{0.5cm}
               & \hspace{0.5cm} \tau  \hspace{0.1cm} }
                                      \hspace{1.0cm} \nonumber \\
N \; = \;
\matrix{ e \hspace{0.2cm} \cr
         \mu \hspace{0.2cm} \cr
         \tau \hspace{0.2cm} }
\left( \matrix{ a  &
                      z+id &
                              y-id \cr
                z-id &
                    b &
                             x+id \cr
      \hspace{2mm} y+id \hspace{2mm} &
         \hspace{2mm}  x-id \hspace{2mm} &
           \hspace{2mm} c  \hspace{2mm} \cr } \right) \label{mmn1}
\end{eqnarray}
where the seven variables $a$, $b$, $c$, $x$, $y$, $z$ and $d\,$ determine 
the three neutrino masses and the four mixing parameters
(the charged-lepton masses being directly the diagonal elements of $L$). 
In terms of these variables the Jarlskog determinant is given by:
\begin{equation}
{\rm Det } \; C \, = \, {\rm Tr} \; C^3/3 \, = \,
2d(m_e-m_{\mu})(m_{\mu}-m_{\tau})(m_{\tau}-m_e) (d^2-xy-yz-zx) \label{cp}
\end{equation}
so that $CP$ violation vanishes in the case $d=0$
and also in the case $d^2=xy+yz+zx$.

The off-diagonal elements of $-i[N,C^2]$ 
must be set to zero for an extremum (Eq.~\ref{exsl}), 
leading to the following cyclically symmetric constraints,
from the real (Re) parts:
\begin{eqnarray}
\hspace{1.0cm}   -d \, (m_e-m_{\mu}) \, (m_{\tau}-m_e) \;
   \left( \rule{0mm}{4mm} \, (y-z) \; - \; (b-c) \, \right) \; (y+z) \;=\; 0 \nonumber \hspace{1.8cm} \\
{\rm Re} \hspace{1.0cm}   -d \, (m_{\mu}-m_{\tau}) \, (m_e-m_{\mu}) \;
   \left( \rule{0mm}{4mm} \, (z-x) \; - \; (c-a) \, \right) \; (z+x) \;=\;0 \hspace{1.8cm} \label{szr} \\ 
\hspace{1.0cm}   -d \, (m_{\tau}-m_e) \, (m_{\mu}-m_{\tau}) \;
   \left( \rule{0mm}{4mm} \, (x-y) \; - \; (a-b) \, \right) \; (x+y) \;=\;0 \nonumber \hspace{1.8cm} 
\end{eqnarray}
and from the imaginary (Im) parts:
\begin{eqnarray}
\hspace{1.0cm}   -(m_e-m_{\mu}) \, (m_{\tau}-m_e) \; 
   \left( \rule{0mm}{4mm} \,(b-c)\,(d^2-yz) \; - \; (y^2-z^2) \, x \right) \;=\;0 \nonumber \hspace{1.4cm} \\
{\rm Im} \hspace{1.0cm}   -(m_{\mu}-m_{\tau}) \, (m_e-m_{\mu}) \; 
   \left( \rule{0mm}{4mm} \,(c-a)\,(d^2-zx) \; - \; (z^2-x^2) \, y \right) \;=\;0 \hspace{1.4cm} \label{szi} \\
\hspace{1.1cm}   -(m_{\tau}-m_e) \, (m_{\mu}-m_{\tau}) \; 
   \left( \rule{0mm}{4mm} \,(a-b)\,(d^2-xy) \; - \; (x^2-y^2) \, z \right) \;=\; 0 . \nonumber \hspace{1.3cm}
\end{eqnarray}
It should be remarked that the cyclic symmetry
($e\to \mu\to \tau$, $a\to b\to c$, $x\to y \to z$)
will be seen to be a useful and important generic feature of our approach, 
resulting from starting with flavour-symmetric invariants
and working in the epsilon basis.

A non-trivial solution to Eq.~\ref{szr},
ie.\ that corresponding to the real parts above
(putting aside the $d=0$ possibility for one moment), may be written: 
\begin{equation}
\matrix{(x-y)=(a-b) \cr (y-z)=(b-c) \cr (z-x)=(c-a) } 
\hspace{1.0cm} {\rm i.e.} \hspace{1.0cm}
\matrix{a=\sigma+x \cr b=\sigma +y \cr c=\sigma+z } \label{s3c}
\end{equation}
where $\sigma$ is an undetermined overall constant offset.
Eq.~\ref{s3c} is precisely the `S3 invariant constraint' \cite{char}
(or `magic square constraint' \cite{simp})  
whereby the neutrino mass matrix is determined to be an `S3 group matrix' \cite{char},
ie.\ a matrix having all row/column sums equal and hence 
(at least) one trimaximal eigenvector.

Turning to Eq.~\ref{szi} (corresponding to the imaginary parts)
a non-trivial solution is:
\begin{eqnarray}
a=b     \hspace{1.5cm} &{\rm and}& \hspace{1.5cm} x=y \nonumber \\
b=c     \hspace{1.5cm} &{\rm and}& \hspace{1.5cm} y=z \label{circ} \\
c=a     \hspace{1.5cm} &{\rm and}& \hspace{1.5cm} z=x \nonumber
\end{eqnarray}
%
whereby the mass matrix is circulant \cite{hs1}
and the mixing takes the trimaximal form \cite{hps1}
(Eq.~\ref{s3c} being also satisfied).
More trivially, coming to the $d=0$ case (above),
we have that twofold-maximal mixing, eg.\ $b=c$, $y=z=0$
also satisfies both sets of equations.

Of course, these solutions
do not solve {\em all} the extremisation equations, Eqs.~\ref{exsl}-\ref{exsn}.
It transpires however, that in the case that the masses are held fixed,
all the remaining equations serve only to determine
suitable Lagrange multipliers $\lambda_{Li}$, $\lambda_{Ni}$ 
($i=0, \, 1,\, 2$):
\begin{eqnarray}
( \partial_L \, {\rm Tr} \; C^3/3 )^T = -i[N,C^2] \;&=& \; 
 \lambda_{L0} \; + \; \lambda_{L1} \, L \; + \; \lambda_{L2} \, L^2 \label{exlml} \\
( \partial_N \, {\rm Tr} \; C^3/3 )^T = +i[L,C^2] \;&=& \;
 \lambda_{N0} \, + \, \lambda_{N1} \, N \, + \, \lambda_{N2} \, N^2
\label{exlmn}
\end{eqnarray} 
with the apparently excess constraints turning out to be redundant
(and furthermore with this circumstance occuring
for all the extremisations studied in this paper).

In the epsilon basis, $L$ is diagonal
and any polynomial in $L$ is diagonal also,
whereby Eqs.~\ref{szr}-\ref{szi} 
(and their solutions, eg.\ Eqs.~\ref{s3c}-\ref{circ}) remain valid,
even with the Lagrange multipliers (Eq.~\ref{exlml}).
In addition, the {\em on}-diagonal elements of $+i[L,C^2]$ 
vanish (as in Eq.~\ref{exsn}) due to $L$ being diagonal,
and in the case of non-zero Lagrange multipliers 
(Eq.~\ref{exlmn}) 
the consistency requirement 
on the coefficients of the $\lambda_{Ni}$ takes the form: 
\begin{equation}
\left| \matrix{ \hspace{2mm} 1 \hspace{2mm} & 
        \hspace{2mm}a \hspace{2mm} & \hspace{2mm}a^2+y^2+z^2\hspace{2mm} \cr
               1 & b & x^2+b^2+z^2 \cr
               1 & c & x^2+y^2+c^2} \right| = 0. \label{detc}
\end{equation}
All our solutions above do indeed turn out to satisfy this determinant condition. 
(In~fact {\em any}~$S3$ group matrix, Eq.~\ref{s3c},
automatically satisfies this condition, Eq.~\ref{detc}).

Covariance implies that the $\lambda_{Li/Ni}$ themselves will be Jarlskog scalars,
expressible in terms of, eg.\ traces of powers of mass matrices:
\begin{eqnarray}
L_1 & := & {\rm Tr} L = m_e+m_{\mu}+m_{\tau} 
              \hspace{1.2cm} N_1 := {\rm Tr} N =  m_1+m_2+m_3 \hspace{1.0cm} \nonumber \\
L_2 & := & {\rm Tr} L^2 = m_e^2+m_{\mu}^2+m_{\tau}^2 
              \hspace{1.0cm} N_2 := {\rm Tr} N^2  =  m_1^2+m_2^2+m_3^2 \hspace{1.0cm} \label{mscons} \\
L_3 & := & {\rm Tr} L^3 = m_e^3+m_{\mu}^3+m_{\tau}^3 
              \hspace{1.0cm} N_3 := {\rm Tr} N^3  =  m_1^3+m_2^3+m_3^3 \hspace{1.0cm} \nonumber
\end{eqnarray}
or traces of commutators of mass matrices etc.
Note that the definitions (Eq.~\ref{mscons}) 
are our `constraint equations', eg.\ ${\rm Tr} \, L^2 \, = \, L_2$ for fixed $L_2$ etc., 
which when differentiated,
$\partial_{L} \, ({\rm Tr } \; L^2 \; - L_2) \, = \, 2L^T$ etc.,
lead to the polynomial forms on the RHS of Eqs.~\ref{exlml}-\ref{exlmn}.

The on-diagonal elements of Eq.~\ref{exlml} 
(which are purely real) are given by:
\begin{eqnarray}
2d \, (m_{\mu}+m_{\tau}-2m_e) \, (m_{\mu}-m_{\tau}) \; (d^2-x^2-y^2-z^2) \; = \; 
\lambda_{L0} \; + \; \lambda_{L1} m_e \; + \; \lambda_{L2} m_e^2    \nonumber \hspace{0.2cm} \\
2d \, (m_{\tau}+m_e-2m_{\mu}) \, (m_{\tau}-m_e) \; (d^2-x^2-y^2-z^2) \; = \;
 \lambda_{L0} \; + \; \lambda_{L1} m_{\mu} \; + \; \lambda_{L2} m_{\mu}^2  \hspace{0.2cm} \label{c311} \\
\hspace{1mm} 
2d \, (m_e+m_{\mu}-2m_{\tau}) \, (m_e-m_{\mu}) \; (d^2-x^2-y^2-z^2) \; = \;
 \lambda_{L0} \; + \; \lambda_{L1} m_{\tau} \; + \; \lambda_{L2} m_{\tau}^2.   \nonumber \hspace{0.1cm}
\end{eqnarray}
Solving 
for the $\lambda_{Li}$ gives 
the charged-lepton discriminant $L_{\Delta}^2$ in the denominator of the solution
(from the determinant-of-coefficients, Eq.~\ref{c311}, RHS),
where $L_{\Delta}$ is given by:
\begin{eqnarray}
L_{\Delta}& :=& \sqrt{L_2^3/2+3L_1^4L_2/2+6L_1L_2L_3-7L_1^2L_2^2/2-3L_3^2-4L_1^3L_3/3-L_1^6/6}
                                                                          \nonumber  \hspace{1.0cm}\\ 
\; & = & \; (m_e-m_{\mu})(m_{\mu}-m_{\tau})(m_{\tau}-m_e). \label{discl}
\end{eqnarray} 
After some work to cast the numerators also into invariant form, we find that: 
\begin{eqnarray}
\lambda_{L0} &=&
 \frac{-L_1^5/2 + 3 L_1^3 L_2 - 7L_1 L_2^2/2 
                     - 2L_1^2 L_3 + 3L_2 L_3}{3L_{\Delta}^2} \; {\rm Tr} \, C^3     \label{c3l0} \\
\lambda_{L1} &=& \frac{3L_1^4/2 - 7 L_1^2 L_2 
                      + 3 L_2^2/2 + 6 L_1 L_3}{L_{\Delta}^2} \; {\rm Tr} \, C^3    \label{c3l1}  \\
\lambda_{L2} &=& \frac {-9L_3 + 9L_1L_2 - 2L_1^3}{3L_{\Delta}^2} \; {\rm Tr} \, C^3. \label{c3l2} 
\end{eqnarray}
Of course, these expressions can also be obtained
in a fully covariant way simply by multiplying Eq.~\ref{exlml}
by successive powers of $L$ ($I=L^0$, $L=L^1$, $L^2$) and taking traces
(powers higher than $L^2$ 
must always be reduced using the characteristic equation).
Analogous $L \leftrightarrow N$ 
expressions for the $\lambda_{Ni}$ solve all the remaining equations.

Recognising the coefficients of ${\rm Tr} \, C^3$ 
on the RHS above 
as $\partial_{L/N} (L_{\Delta}N_{\Delta})/(L_{\Delta}N_{\Delta})$, 
we see that we have, in effect, extremised 
the Jarlskog invariant $J={\rm Tr} C^3/(L_{\Delta}N_{\Delta})$,
recovering the known result \cite{sqt3}, and validating our procedure.
Finally, we note that the above analysis is essentially unchanged,
taking $L$ and $N$ as the hermitian squares of mass matrices (see Section~1)
substituting masses-squared for masses throughout. \\

\noindent {\bf 3. The Sum of the 
\begin{boldmath} $2 \times 2$ \end{boldmath}
 Principal Minors,} 
\begin{boldmath}${-\rm Tr} \, C^2/2$\end{boldmath}
\vspace{2mm}

\noindent 
The other independent invariant function of $C$
may be taken to be the sum of the $2 \times 2$ principal minors, 
expressible in terms of the $K$-matrix ($K_{l\nu} :=-K_{l' l''}^{\nu'\nu''} $\cite{kmat}) as follows:
\begin{equation}
Q_{11}= -{\rm Tr} \, C^2/2= {\rm Tr} \, [L,N]^2/2=
\Delta_l^T \; {\rm diag}(\Delta_l) 
                      \; K \; 
                          {\rm diag}(\Delta_{\nu}) \; \Delta_{\nu}. 
\label{trc2}
\end{equation}
The $K$-matrix is a key oscillation observable \cite{kmat},
carrying equivalent information
to the moduli-squared of the mixing elements \cite{box}. 
The $K$-matrix comprises the real parts of the plaquette products \cite{bjpi}: 
$\Pi_{l\nu} :=-K_{l\nu} + iJ$ (indices to be interpreted mod~3) 
and may be viewed as the $CP$-conserving analogue 
of the Jarlskog invariant $J$ \cite{jcp}. 

From Eq.~\ref{trc2}, taking for simplicity initially 
the extreme hierarchical approximation
$m_e$, $m_{\mu}$, $m_{1}$, $m_{2}$ $\rightarrow 0$
(ie.\ the $2 \times 2$ mixing limit), we have:
\begin{eqnarray}
Q_{11}  
\rightarrow m_{\tau}^2m_3^2 \; (K_{e 1}+K_{e 2}+K_{\mu 1} + K_{\mu 2})
                                  + O(m_{\tau}^2 m_2^2) +  \dots \nonumber \\
= m_{\tau}^2m_3^2 \; c_{13}^2c_{23}^2(s_{13}^2+c_{13}^2s_{23}^2) 
                           + O(m_{\tau}^2 m_2^2) + \dots \hspace{1.2cm} \label{kpdg}
\end{eqnarray} 
Extremising this expression with respect to $s_{23}$ and $s_{13}$ gives
$s_{23}=s_{13}=0$ or $s_{23}=s_{13}=1$ or
$c_{23} c_{13}=1/ \sqrt{2}$ 
(where the latter looks at first sight very promising phenomenologically, see below).
Extremising rather with respect to $\theta_{23}$ and $\theta_{13}$
gives in addition $s_{23}=1$ or $s_{13}=1$ as solutions,
for arbitrary $s_{13}$ and $s_{23}$ respectively.
Since mixing matrices which are just permutation matrices 
give no mixing as such,
we have that $2 \times 2$ maximal mixing 
is the only non-zero mixing solution in the $2 \times 2$ case.

Just as for Tr $C^3$ above, however, we will extremise exactly here with respect to 
the mass matrices (see again Appendix~A):
\begin{eqnarray}
(-\partial_L {\rm Tr} \, C^2/2)^T = +[N,[L,N]] & = & \; 0, \hspace{3mm}
 \lambda_{L0} \; + \; \lambda_{L1} \, L \; + \; \lambda_{L2} \, L^2 \label{expl} \\
(-\partial_N {\rm Tr} \, C^2/2)^T = -[L,[L,N]] & = & \; 0, \hspace{3mm}
 \lambda_{N0} \, + \, \lambda_{N1} \, N \, + \, \lambda_{N2} \, N^2 \label{expn}
\end{eqnarray}  
where the matrix polynomials replace the RHS zeros, 
in the case of mass constraints.
In the epsilon basis the off-diagonal elements 
of 
Eq.~\ref{expl} are,
for the real parts:
%
\begin{eqnarray}
 \hspace{2.0cm}  \; (m_{\mu}+m_{\tau}-2m_e) \, (d^2-y \, z) \; +
         \;  (m_{\mu}-m_{\tau}) \, (b-c) \, x \; = \; 0 \hspace{1.6cm} \nonumber \\
{\rm Re} \hspace{1.0cm}  (m_{\tau}+m_e-2m_{\mu}) \, (d^2-z \, x) \; +
         \;  (m_{\tau}-m_e) \, (c-a) \, y \; = \; 0 \hspace{1.6cm} \label{rzr}  \\
 \hspace{2.0cm}  (m_e+m_{\mu}-2m_{\tau}) \, (d^2-x \, y) \; +
         \;  (m_e-m_{\mu}) \, (a-b) \, z \; = \; 0 \hspace{1.6cm} \nonumber
\end{eqnarray}
and for the imaginary parts:
\begin{eqnarray}
 \hspace{2.0cm}  d \, (m_{\mu}+m_{\tau}-2m_e) \, (y+z) \; + 
          \;  d \, (m_{\mu}-m_{\tau}) \, (b-c) \; = \; 0 \hspace{1.6cm} \nonumber \\
{\rm Im}  \hspace{1.0cm}  d \, (m_{\tau}+m_e-2m_{\mu}) \, (z+x) \; + 
          \;  d \, (m_{\tau}-m_e) \, (c-a) \; = \; 0 \hspace{1.6cm} \label{rzi} \\
\hspace{2.0cm}  d \, (m_e+m_{\mu}-2m_{\tau}) \, (x+y) \; + 
          \;  d \, (m_e-m_{\mu}) \, (a-b) \; = \; 0. \hspace{1.6cm} \nonumber 
\end{eqnarray}
We observe that Eqs.~\ref{rzr}-\ref{rzi} 
imply that either $d=0$ or $d^2=xy+yz+zx$, 
and so from Eq.~\ref{cp} we see immediately that 
there can be no $CP$ violation in this case.

It turns out that the $d^2=xy+yz+zx$ solution is equivalent
to the $d=0$ solution, so that we need to consider only the $d=0$ case 
(note that in the case $d^2=xy+yz+zx$ the epsilon basis
is not unique, and the imaginary part can be rephased to zero).   
Setting $d=0$ then,
solves Eq.~\ref{rzi}, corresponding to the imaginary parts.
Clearly, setting any pair of $x$, $y$, $z$ to zero
and the corresponding pair of $a$, $b$, $c$ equal,
solves also Eq.~\ref{rzr} for the real parts,
so that, just as for Tr $C^3$ above, $2 \times 2$ maximal mixing in any sector
gives an exact extremum of Tr $C^2$, 
independent of the neutrino mass spectrum.

A somewhat more interesting non-trivial solution
to Eq.~\ref{rzr} (still with $d=0$ from Eq.~\ref{rzi}) 
clearly reflects the inherent cyclic symmetry:
%
\begin{eqnarray}
x=\pm \sqrt{(a-b)(c-a)E} \hspace{1.0cm} 
E=\frac{(m_e-m_{\mu})(m_{\tau}-m_e)}{(m_e+m_{\mu}-2m_{\tau})(m_{\tau}+m_e-2m_{\mu})} \nonumber \\
y=\pm \sqrt{(b-c)(a-b)M} \hspace{1.0cm} 
M=\frac{(m_{\mu}-m_{\tau})(m_e-m_{\mu})}{(m_{\mu}+m_{\tau}-2m_e)(m_e+m_{\mu}-2m_{\tau})} \label{solr} \\
z=\pm \sqrt{(c-a)(b-c)T} \hspace{1.0cm} 
T=\frac{(m_{\tau}-m_e)(m_{\mu}-m_{\tau})}{(m_{\tau}+m_e-2m_{\mu})(m_{\mu}+m_{\tau}-2m_e)}. \nonumber
\end{eqnarray}
With the charged-lepton masses positive, we have $E > 0 > M,T$,
whereby either $\, b \, < a \, < \, c \; $ or $\; c \, < a \, < \, b \, $.
We may take all upper (positive) signs 
in the case $b \, < a \, < \, c$,
corresponding to an inverted hierarchy,
and all lower (negative) signs in the case $c \, < a \, < \, b$,
corresponding to a normal hierarchy (as considered below)
with other sign combinations 
being just trivial rephasings of these two possibilities.
For these solutions the determinant condition
is again automatically satisfied, since $E+M+T+1=0$
appears as a factor in the determinant Eq.~\ref{detc}. 

As in the case of ${\rm Tr} \, C^3$ above,
the remaining constraints (from Eqs.~\ref{expl}-\ref{expn}) 
will serve only to fix the Lagrange multipliers.
While we again have the option to determine the Lagrange multiplers 
in a general and fully covariant way,
it will prove useful as before 
to consider eg.\ the on-diagonal $\partial_L$ equations 
(which are again purely real) explicitly:
\begin{eqnarray}
\hspace{0.12cm}   2(m_{\tau}-m_e) y^2 \, - \, 2(m_e-m_{\mu}) z^2 & = & 
\lambda_{L0} \; + \; \lambda_{L1} \, m_e \; + \; \lambda_{L2} \, m_e^2 
                                                  \nonumber \hspace{0.27cm} \\
\hspace{0.2cm}  2(m_e-m_{\mu})z^2 \, - \, 2(m_{\mu}-m_{\tau})x^2 & = &
 \lambda_{L0} \; + \; \lambda_{L1} \, m_{\mu} \; + \; \lambda_{L2} \, m_{\mu}^2 
                                                \hspace{0.2cm} \label{c211} \\
\hspace{0.2cm}   2(m_{\mu}-m_{\tau})x^2 \, - \, 2(m_{\tau}-m_e)y^2 & = &
 \lambda_{L0} \; + \; \lambda_{L1} \, m_{\tau} \; + \; \lambda_{L2} \, m_{\tau}^2. 
                                                \nonumber \hspace{0.2cm}
\end{eqnarray}
Since Lagrange multipliers need in general 
only be evaluated locally at the extremum,
we now have the possibility to achieve
some simplification by substituting 
(on the LHS of Eq.~\ref{c211})
the specific solutions Eq.~\ref{solr},
before solving for the $\lambda_{Li}$ 
and again casting the resulting expressions 
(curly brackets denote anticommutators)
in invariant form: 
\begin{eqnarray}
\lambda_{L0} & = &
 \frac{({\rm Tr} \, \{ L^2,N \} \, L_1 \; - \; {\rm Tr} \, \{ L,N \} \, L_2)
               (3{\rm Tr} \, \{ L,N \} \; - \; 2L_1 N_1)}{2(9L_3-9L_1L_2+2L_1^3)} \label{c2lm0} \\
\lambda_{L1} & = & \frac{-(3{\rm Tr} \, \{ L^2,N \} - 2L_2N_1)
               (3{\rm Tr} \, \{ L,N \} \; - \; 2L_1 N_1)}{2(9L_3-9L_1L_2+2L_1^3)} \label{c2lm1} \\
\lambda_{L2} & = & \frac {(3{\rm Tr} \, \{ L,N \} \; - \; 2L_1 N_1)^2}{2(9L_3-9L_1L_2+2L_1^3)}.  
\label{c2lm2}
\end{eqnarray}
(Similar expressions also result for $L \leftrightarrow N$, as before).
While this 
means that our Lagrange multipliers are now solution-specific
(ie.\ they fail for the mass-independent 
$2 \times 2$ maximal-mixing solution above),
one at least has that Eqs.~\ref{expl}-\ref{expn} 
and Eqs.~\ref{c2lm0}-\ref{c2lm2} 
(with their $L \leftrightarrow N$ counterparts)
together constitute a covariant statement of specifically
the non-trivial (cyclically symmetric) solution 
given by Eq.~\ref{solr}.

From Eq.~\ref{solr} the mixing is uniquely predicted, 
in essence by fitting $a$, $b$ and $c$ (Eq.~\ref{mmn1})
to give the correct neutrino mass spectrum. 
Of course absolute neutrino masses
(as distinct from mass-squared differences)
are not yet known directly, 
but we may still obtain a specific prediction
if we are prepared to assume 
a normal (ie.\ non-inverted) 
`classic' fermion mass hierarchy  $m_1 << m_2 << m_3$
(ie. with no offset, $m_1 \simeq 0$).

We then find that
an acceptable neutrino mass-squared difference hierarchy 
($h_{\nu}^2 \, :=\Delta m_{12}^2/\Delta m_{23}^2 \simeq m_2^2/m_3^2 \simeq 0.03$)
results, simply by setting the ratio $(b-a)/(a-c) = 1$,
corresponding, it turns out, to the $\nu_e$ being exactly trimaximally mixed
(this ratio is in fact the only operative parameter here,
since rescaling and off-setting of $a$, $b$ and $c$ 
clearly does not influence the mixing matrix).
The resulting $|U|^2$-matrix (ie.\ the matrix of 
moduli-squared
of the resulting mixing-matrix elements) is given numerically by:
\begin{eqnarray}
     \matrix{  \hspace{0.1cm} \nu_1 \hspace{0.6cm}
               & \hspace{0.4cm} \nu_2 \hspace{0.6cm}
               & \hspace{0.4cm} \nu_3  \hspace{0.2cm} }
                                      \hspace{2.4cm} 
          \matrix{  \hspace{0.1cm} \nu_1 \hspace{0.2cm}
               & \hspace{0.4cm} \nu_2 \hspace{0.2cm}
               & \hspace{0.4cm} \nu_3  \hspace{0.4cm} }
                                      \hspace{0.4cm}\nonumber \\
(|U_{l \nu}|^2) \; = \;
\matrix{ e \hspace{0.2cm} \cr
         \mu \hspace{0.2cm} \cr
         \tau \hspace{0.0cm} } \hspace{-2mm}
\left( \matrix{ .33333  &
                      .33333 &
                              .33333 \cr
                .17079 &
                    .16257 &
                             .66663  \cr
      \hspace{2mm} .49587 \hspace{2mm} &
         \hspace{2mm}  .50409 \hspace{2mm} &
           \hspace{2mm} .00003  \hspace{2mm} \cr } \right)
\; \simeq \;
\matrix{ e \hspace{0.2cm} \cr
         \mu \hspace{0.2cm} \cr
         \tau \hspace{0.0cm} } \hspace{-2mm}
\left( \matrix{ 1/3  &
                      1/3 &
                              1/3 \cr
                1/6 &
                    1/6 &
                             2/3  \cr
      \hspace{2mm} 1/2 \hspace{2mm} &
         \hspace{2mm}  1/2 \hspace{2mm} &
           \hspace{2mm} 0  \hspace{2mm} \cr } \right). \hspace{2mm} \label{akeq}
\end{eqnarray}
While being 
closely a transpose/permutation of the tri-bimaximal form,
and bearing a clear relation to some previously proposed ansatze
(see in particular Acker et al.\ \cite{ack} and Karl et al. \cite{karl}
as well as the Fritzsch-Xing ansatz \cite{fxng}),
this mixing (Eq.~\ref{akeq}) clearly cannot at this point 
be considered acceptable phenomenologically.

We find that the situation is not improved 
substituting masses-squared in place of masses throughout. 
Indeed the large discrepancy in $|U_{\tau 3}|$
is considerably worsened,
with $|U_{\tau3}|^2 \sim 3 \times 10^{-9}$
(maintaining the mass hierarchy unchanged 
requires setting $(b-a)/(a-c) \sim 0.36$).
Having also explored less minimalist assumptions
regarding the neutrino mass-spectrum, 
ie.\ inverted mass hierarchy, large mass offsets etc.\
we can report no phenomenlogically acceptable solutions
to the problem of extremising Tr $C^2$.
For $m_1 << m_2 << m_3$, we see that  
all our exact solutions are consistent
with the (approximate) constraints
derived differentiating Eq.~\ref{kpdg} with respect to PDG angles
(although we learn that the initially promising $U_{\tau 3}=c_{23}c_{13} =1/\sqrt{2}$ condition 
in fact
corresponds only to phenomenologically uninteresting $2 \times 2$ maximal mixing solutions). \\

\noindent {\bf 4. Extremising More General Mixing Invariants }  
\vspace{2mm}

\noindent
More general commutator invariants are readily constructed,
involving higher powers of mass matrices \cite{bran}.
In this section we shall extremise a very simple composite function,
comprising an arbitrary linear combination of two such invariants:
\begin{equation}
A \, := \, Q_{11}+qQ_{21} \label{actrq}
\end{equation}
with $q$ a scalar constant of suitable dimensionality (assumed Jarlskog-invariant) and:
\begin{eqnarray}
Q_{11} := {\rm Tr} \; [L,N] \, [L,N] \;/ \, 2 \hspace{1.0cm} 
Q_{21} := {\rm Tr} \; [L,N] \, [L^2,N] \; / \, 2,   \label{pq}  
\end{eqnarray}
our $Q_{mn}$ notation being essentially self-explanatory 
\footnote{
We are considering here flavour-symmetric
quadratic commutator invariants 
(vanishing in the case of zero mixing)
which may be usefully arranged as a $3 \times 3$ matrix:
\begin{equation}
Q \; = \; \frac{1}{2}
          \left( \matrix { {\rm Tr} \, [L,N]^2  & {\rm Tr} \, [L,N][L,N^2] & {\rm Tr} \, [L,N^2]^2  \cr
                 {\rm Tr} \, [L,N][L^2,N] &  {\rm Tr} \, [L,N][L^2,N^2] & {\rm Tr} \, [L,N^2][L^2,N^2]  \cr
                 {\rm Tr} \, [L^2,N]^2 & {\rm Tr} \, [L^2,N][L^2,N^2] &  {\rm Tr} \, [L^2,N^2]^2 } \right).
\label{qmat}
\end{equation}
As usual, powers of mass matrices higher 
than $L^2$, $N^2$ are not considered,
by virtue of the characteristic equation
(and ${\rm Tr} \; [L^2,N] \, [L,N^2] \equiv {\rm Tr} \; [L,N] \, [L^2,N^2]$,
so there are indeed only nine such invariants).  
Generalising Eq.~\ref{trc2}, the $Q_{mn}$ ($m$,$n$ $=$ $1,2,3$)
are the double moments of the K-matrix:
\begin{eqnarray}
Q_{mn}=\Delta_l^T \; {\rm diag}(\Delta_l) \;
                     ({\rm diag} \, \Sigma_l)^{m-1}                                    
                      \; K \;   ({\rm diag} \, \Sigma_{\nu})^{n-1} \;
                          {\rm diag}(\Delta_{\nu}) \; \Delta_{\nu} \hspace{14mm} \\
\Sigma_l=(m_{\mu}+m_{\tau},\, m_{\tau}+m_{e}, \, m_{e}+m_{\mu}) \hspace{0.5cm}
\Sigma_{\nu}=(m_{2}+m_{3},\, m_{3}+m_{1}, \, m_{1}+m_{2}).
\end{eqnarray}
For known masses, the $Q$-matrix and the $K$-matrix
are therefore equivalent,
and both are equivalent to the $|U_{l\nu}|^2$-matrix (see \cite{kmat} \cite{box}),
with only four elements functionally independent in each case. }
at this point.

Of course, $Q_{11}$   
has already been individually extremised in Section~3 (Eqs.~\ref{expl}-\ref{expn}),
so we only need in addition here
the relevant derivatives of $Q_{21}$ (see Appendix~A):
\begin{eqnarray}
(\partial Q_{21} / \partial L)^T & = & [N,[L^2,N]]/2 \, + \, \{ L,[N,[L,N]] \}/2  \label{delq21} \\
(\partial Q_{21} / \partial N)^T & = & -[L^2,[L,N]]                               \label{denq21}
\end{eqnarray}
where 
again curly brackets signify the anti-commutator.

The off-diagonal terms derived from $Q_{21}$, 
analogously to Eq.~\ref{rzr}, are (Re parts):
\begin{eqnarray}
(m_{\mu}^2+m_{\tau}^2+m_{\mu}m_{\tau}-m_{\tau}m_e-m_em_{\mu}-m_e^2) \, (d^2-y \, z) \; + 
                   \; (m_{\mu}^2-m_{\tau}^2) \, (b-c) \, x \hspace{0.1cm} \nonumber \\ 
(m_{\tau}^2+m_e^2+m_{\tau}m_e-m_em_{\mu}-m_{\mu}m_{\tau}-m_{\mu}^2) \, (d^2-z \, x) \; + 
                   \; (m_{\tau}^2-m_e^2) \, (c-a) \, y \hspace{0.1cm}\\
(m_e^2+m_{\mu}^2+m_em_{\mu}-m_{\mu}m_{\tau}-m_{\tau}m_e-m_{\tau}^2) \, (d^2-x \, y) \; + 
                   \; (m_e^2-m_{\mu}^2) \, (a-b) \, z \hspace{0.1cm} \nonumber
\end{eqnarray}
and (Im parts):
\begin{eqnarray}
d \, (m_{\mu}^2+m_{\tau}^2+m_{\mu}m_{\tau}-m_{\tau}m_e-m_em_{\mu}-m_e^2) \, (y+z) \; + 
                     \; d \, (m_{\mu}^2-m_{\tau}^2) \, (b-c) \hspace{0.1cm} \nonumber \\
d \, (m_{\tau}^2+m_e^2+m_{\tau}m_e-m_em_{\mu}-m_{\mu}m_{\tau}-m_{\mu}^2) \, (z+x) \; + 
                     \; d \, (m_{\tau}^2-m_e^2) \, (c-a) \hspace{0.1cm} \\
d \, (m_e^2+m_{\mu}^2+m_em_{\mu}-m_{\mu}m_{\tau}-m_{\tau}m_e-m_{\tau}^2) \, (x+y) \; + 
                     \; d \, (m_e^2-m_{\mu}^2) \, (a-b) \hspace{0.1cm} \nonumber
\end{eqnarray}
and the combined constraint equations are obtained
simply by adding these terms (weighted by $q$) 
into Eqs.~\ref{rzr}-\ref{rzi}.

As in Section~3 we may again set $d=0$, 
and the solutions to these combined equations 
are then readily written down from Eqs.~\ref{solr},
by modifying the expressions for $E$, $M$ and $T$
in Eq.~\ref{solr}, making the (cyclically-symmetric) substitutions:
\begin{eqnarray}
(m_e-m_{\mu}) & \rightarrow & (m_e-m_{\mu})+q(m_e^2-m_{\mu}^2)  \nonumber \\
(m_{\mu}-m_{\tau}) & \rightarrow & (m_{\mu}-m_{\tau})+q(m_{\mu}^2-m_{\tau}^2) \label{subb} \\
(m_{\tau}-m_e) & \rightarrow & (m_{\tau}-m_e)+q(m_{\tau}^2-m_e^2)  \nonumber \\
(m_e+m_{\mu}-2m_{\tau}) & \rightarrow & (m_e+m_{\mu}-2m_{\tau})
                                 +q(m_e^2+m_{\mu}^2+m_em_{\mu}-m_{\tau}L_1) \nonumber \\
(m_{\mu}+m_{\tau}-2m_e) & \rightarrow & (m_{\mu}+m_{\tau}-2m_e)
                                 +q(m_{\mu}^2+m_{\tau}^2+m_{\mu}m_{\tau}-m_eL_1) \label{suba} \\
(m_{\tau}+m_e-2m_{\mu}) & \rightarrow & (m_{\tau}+m_e-2m_{\mu})
                                 +q(m_{\tau}^2+m_e^2+m_{\tau}m_e-m_{\mu}L_1). \nonumber
\end{eqnarray}
The determinant condition Eq.~\ref{detc} 
remains valid since the equation $E+M+T+1=0$
(see text just after Eq.~\ref{solr})
itself remains valid under these substitutions.

Fitting to the `mass' spectrum as in Section~3,
we now have (replacing the unique mixing prediction Eq.~\ref{akeq})
a trajectory of mixings, 
each mixing given by a different value of the parameter $q$.
To locate the phenomenologically viable solution
along this trajectory (supposing even that there is a viable solution),
we will simply impose the ``$S3$ constraint'' (Eq.~\ref{s3c}),
which although not forced on us here by the extremisation,
is well-known \cite{simp} \cite{char} to be consistent with the phenomenology.
We shall see below how the $S3$-constraint (by its nature) 
can be imposed in a fully covariant way.

In practice, we shall substitute 
the $S3$ constraint Eq.~\ref{s3c} into the solution Eq.~\ref{solr} 
(as modified by the substitutions Eqs.~\ref{subb}-\ref{suba}),
thereby fixing all the 
parameters in the neutrino mass matrix, 
as functions of $q$, as follows:
\begin{eqnarray}
a & = & x \, + \, \sigma \hspace{1.0cm} x = \frac{k^3}
                            {(m_{\mu} - m_{\tau})^2(1 + q(m_{\mu} + m_{\tau}))} \nonumber \\
b & = & y \, + \, \sigma \hspace{1.0cm} y = \frac{k^3}
                            {(m_{\tau} - m_e)^2(1 + q(m_{\tau} + m_e))} \label{rqsols3} \\
c & = & z \, + \, \sigma \hspace{1.0cm} z = \frac{k^3}
                            {(m_e - m_{\mu})^2(1 + q(m_e + m_{\mu}))} \nonumber
\end{eqnarray}
where the mixing clearly cannot depend on the scale factor $k$
or the offset $\sigma$.

Assuming a normal classic neutrino mass hierarchy as before,
the operative parameter is now $q$,
fixing both the hierarchy ratio $h_{\nu}^2 \, :=\Delta m_{12}^2/\Delta m_{23}^2 \simeq m_2^2/m_3^2 $
and the mixing simultaneously.  
We see immediately from Eq.~\ref{rqsols3} that,
eg.\ as $q \rightarrow -1/(m_{\mu}+m_{\tau})$ 
we have a pole where $x \rightarrow -\infty$
while the hierarchy factor $h_{\nu} \rightarrow 0$,
with the mixing approaching the exact tri-bimaximal form \cite{tbm1} \cite{tbm2} in the limit.

Parametrising deviations from this pole 
in the form $q=-(1+\epsilon)/(m_{\mu}+m_{\tau})$,  
we find: $\epsilon \simeq 4/3 \; h_l^2h_{\nu}(1+h_{\nu}/3+8h_{\nu}^2/9 + \dots)$.   
For $h_{\nu}^2 \simeq 0.03$ (with $h_l \, := \, m_{\mu}/m_{\tau} \simeq 0.06$)
we have $\epsilon \simeq (0.03)^2$, 
and taking $k \simeq 0.38$ meV$^{1/3}$GeV$^{2/3}$
and $\sigma \simeq 25$ meV,
gives a neutrino mass spectrum 
($m_1$, $m_2$, $m_3$) $\simeq$ ($0$, $8.7$, $50$) meV, 
and a mixing matrix:
\begin{eqnarray}
     \matrix{  \hspace{0.1cm} \nu_1 \hspace{0.6cm}
               & \hspace{0.4cm} \nu_2 \hspace{0.6cm}
               & \hspace{0.4cm} \nu_3  \hspace{0.1cm} }
                                      \hspace{2.6cm} 
          \matrix{  \hspace{0.1cm} \nu_1 \hspace{0.2cm}
               & \hspace{0.4cm} \nu_2 \hspace{0.2cm}
               & \hspace{0.4cm} \nu_3  \hspace{0.2cm} }
                                      \hspace{0.8cm}\nonumber \\
(|U_{l \nu}|^2) \; = \;
\matrix{ e \hspace{0.2cm} \cr
         \mu \hspace{0.2cm} \cr
         \tau \hspace{0.0cm} } \hspace{-2mm}
\left( \matrix{ 0.662  &
                      0.333 &
                              0.005 \cr
                0.219 &
                      0.333 &
                             0.448 \cr
   \hspace{2mm} 0.120 \hspace{2mm} &
         \hspace{2mm}  0.333 \hspace{2mm} &
           \hspace{2mm} 0.547  \hspace{2mm} \cr } \right)
\; \simeq \;
\matrix{ e \hspace{0.2cm} \cr
         \mu \hspace{0.2cm} \cr
         \tau \hspace{0.0cm} } \hspace{-2mm}
\left( \matrix{ 2/3  &
                      1/3 &
                              0 \cr
                1/6 &
                    1/3 &
                             1/2  \cr
      \hspace{2mm} 1/6 \hspace{2mm} &
         \hspace{2mm}  1/3 \hspace{2mm} &
           \hspace{2mm} 1/2  \hspace{2mm} \cr } \right). \label{rqmix} \hspace{2mm}
\end{eqnarray}   
The mixing Eq.~\ref{rqmix}, being very close to the tribimaximal form, 
is certainly acceptable phenomenologically.
There is the non-trivial prediction $|U_{e3}| \simeq \sqrt{2}/3 \; h_{\nu}(1-h_l)^2 \simeq 0.07$
(cf.\ Ref.~\cite{simp}) 
and violations 
of $\mu-\tau$ symmetry \cite{mutau} also, 
eg.\  $\theta_{23} \simeq \pi/4-|U_{e3}|/\sqrt{2}$.
Related poles in $y$ and $z$,
for which the mixing matrix 
would have rows correspondingly permuted
with respect to Eq.~\ref{rqmix},
would seem not to be phenomenologically relevant.

Repeating the analysis with masses-squared,
in effect takes us closer to the pole
(now $\epsilon \sim 5 \times 10^{-7}$)
yielding a mixing correspondingly closer to the tri-bimaximal form:
\begin{eqnarray}
     \matrix{  \hspace{0.1cm} \nu_1 \hspace{0.6cm}
               & \hspace{0.4cm} \nu_2 \hspace{0.6cm}
               & \hspace{0.4cm} \nu_3  \hspace{0.1cm} }
                                      \hspace{2.6cm} 
          \matrix{  \hspace{0.1cm} \nu_1 \hspace{0.2cm}
               & \hspace{0.4cm} \nu_2 \hspace{0.2cm}
               & \hspace{0.4cm} \nu_3  \hspace{0.2cm} }
                                      \hspace{0.8cm}\nonumber \\
(|U_{l \nu}|^2) \; = \;
\matrix{ e \hspace{0.2cm} \cr
         \mu \hspace{0.2cm} \cr
         \tau \hspace{0.0cm} } \hspace{-2mm}
\left( \matrix{ 0.6665  &
                      0.3333 &
                              0.0002 \cr
                0.1768 &
                      0.3333 &
                             0.4898 \cr
   \hspace{2mm} 0.1567 \hspace{2mm} &
         \hspace{2mm}  0.3333 \hspace{2mm} &
           \hspace{2mm} 0.5100  \hspace{2mm} \cr } \right)
\; \simeq \;
\matrix{ e \hspace{0.2cm} \cr
         \mu \hspace{0.2cm} \cr
         \tau \hspace{0.0cm} } \hspace{-2mm}
\left( \matrix{ 2/3  &
                      1/3 &
                              0 \cr
                1/6 &
                    1/3 &
                             1/2  \cr
      \hspace{2mm} 1/6 \hspace{2mm} &
         \hspace{2mm}  1/3 \hspace{2mm} &
           \hspace{2mm} 1/2  \hspace{2mm} \cr } \right). \label{rqmix2} \hspace{2mm}
\end{eqnarray}   
The corresponding prediction for $|U_{e3}|$ 
(now $|U_{e3}| \sim \sqrt{2}/3 \; h_{\nu}^2 \sim 0.014$)
is clearly of less immediate phenomenological interest 
than that in the unsquared case above.

Lagrange multipliers for Eq.~\ref{actrq} are found in the usual way.
For the $\partial_L$ equations,
solving the analogue of Eq.~\ref{c211} (after the substitutions Eq.~\ref{subb}),
we find:
\begin{eqnarray}
\lambda_{L0} = N_1^2\frac{\lambda_{L00} + \lambda_{L01}q + \lambda_{L02}q^2 + 
               \lambda_{L03}q^3 + \lambda_{L04}q^4 + \lambda_{L05}q^5}
             {2(q^2L_{P6} + 2qL_{P2}L_{P3} + L_{P2}^2)^2}  \label{rqlm0l} \\
\lambda_{L1} = N_1^2\frac{\lambda_{L10} + \lambda_{L11}q + \lambda_{L12}q^2 + 
               \lambda_{L13}q^3 + \lambda_{L14}q^4 + \lambda_{L15}q^5}
             {2(q^2L_{P6} + 2qL_{P2}L_{P3} + L_{P2}^2)^2} \\
\lambda_{L2} = N_1^2\frac{\lambda_{L20} + \lambda_{L21}q + \lambda_{L22}q^2 + 
               \lambda_{L23}q^3 + \lambda_{L24}q^4 + \lambda_{L25}q^5}
             {2(q^2L_{P6} + 2qL_{P2}L_{P3} + L_{P2}^2)^2}
\end{eqnarray}
where we have explicitly assumed the `$S3$ constrained' solution Eq.~\ref{rqsols3}.
Fully invariant and flavour-symmetric,
the expressions for $\lambda_{L00}$, $\lambda_{L01}$ \dots $\lambda_{L10}$ etc.,
as functions of the $L_i$, $i=1 \dots 3$, are given in Appendix~B.

Similarly, for the $\partial_N$ equations we find:
\begin{eqnarray}
\lambda_{N0} & = & -N_2\frac{L_{P2}^2 + 2qL_{P2}L_{P3} + q^2L_{P6}}{6N_1(L_{P2} + qL_{P3})} \\
\lambda_{N1} & = & 0 \\
\lambda_{N2} & = & \frac{L_{P2}^2 + 2qL_{P2}L_{P3} + q^2L_{P6}}{2N_1(L_{P2} + qL_{P3})} \label{rqlm2n}
\end{eqnarray} 
with no clear relation to the $\partial_L$ case, 
since our ``action'' (Eq.~\ref{actrq}) is not $L \leftrightarrow N$ sym- metric.
The supplementary polynomials $L_{P2}$, $L_{P3}$, $L_{P6}$ are defined in Appendix B.

It should be emphasised that 
(up to the apparently spontaneous choice of pole) 
the extremisation conditions on the composite action Eq.~\ref{actrq},
together with the particular Lagrange multipliers 
defined by Eqs.~\ref{rqlm0l}-\ref{rqlm2n} and Appendix~B, 
constitute an entirely covariant and flavour-symmetric
specification of the realistic, non-trivial mixing Eq.~\ref{rqmix}. \\

\noindent {\bf 5. Perspective }  
\vspace{2mm}

\noindent
Motivated by the notion (Section~1) 
that the fundamental laws 
underlying the fermion masses and mixings
might come from a variational principle,
applied to some as yet unspecified function
of flavour-symmetric Jarlskog invariants,
we have presented a manifestly covariant machinary
for extremising such invariants as functions of the mass matrices.
(Building on the analogy \cite{hs1} 
with the field-strength tensor 
in terms of covariant derivatives, 
cf.\ $F_{\mu \nu}=-i[{\cal D}_{\mu},{\cal D}_{\nu}]$,
we see our extremisation equations,
eg.\ Eqs.~\ref{expl}-\ref{expn}
as analogous to the Yang-Mills \cite{yngm} equations,
cf.\ $[{\cal D}_{\mu}, F^{\mu \nu}]=0$,
which are themselves derivable from 
a quadratic Lagrangian ${\cal L} = -{\rm Tr} \; F^2/2$,
whereby we see the Yukawa couplings as ``dynamical variables'',
somewhat analogous to the gauge fields).

Having tested-out our methods 
on the familiar case of Tr $C^3$ (Section~2) 
we have been able to show (Section~3) 
that extremising Tr $C^2$ does 
not lead to physically realistic lepton mixings. 
Focussing on quadratic ``actions'', 
we see Eq.~\ref{actrq} 
(Section~4) as just an example
of an (approximate) ``effective'' action, 
whose main particular merit is that we know how to solve 
its resulting extremisation equations analytically,
and that it leads to at least one mixing (Eq.~\ref{rqmix})
which is compatable with observation.
Clearly the complete absence of $CP$ violation
is a significant deficiency of Eq.~\ref{actrq} (given the sucesss
of the standard explanation of $CP$ violation in the quark sector)
and incorporating $CP$ violation 
is an obvious challenge for the future.

Most of all, 
we would like to think that this paper 
will stimulate the search for
a natural and beautiful action function,
determining not only the mixing 
but also the (relative) mass values 
$m_1 : m_2 : m_3$ and
$m_e : m_{\mu} : m_{\tau}$,
obviating the need for those ugly Lagrange multipliers 
(and perhaps also shedding some light 
on the quark case at the same time).
Regarding masses, it is clear 
that our procedures apply directly as they stand:
one will only have to notice 
that for some `perfect' action,
not only is the mixing correctly predicted,
but the Lagrange multiplier functions, 
on the RHS of the extremisation equations,
vanish numerically for the correct masses,
effectively as in a free extremisation.
(Empirical relations like the Koide relation:
$512L_3L_1 - 64L_2^2 - 656L_2L_1^2 + 207L_1^4=0$ \cite{koid}
could  be relevant here, cf.\ Appendix~B).
In removing a large part of the arbitrariness
inevitably associated with the flavour sector as we know it today,
such an ``action'' (if it exists) would surely be welcomed
as an economical adjunct to the present Standard Model of Particle Physics. \\

\noindent {\bf Acknowledgements }  
\vspace{1mm}

\noindent
It is a pleasure to thank S.\ Pakvasa 
and T.\ J.\ Weiler for useful discussions 
and helpful comments. 
This work was supported by the UK 
Particle Physics and Astronomy Research Council (PPARC).
One of us (PFH) acknowledges the hospitality 
of the Centre for Fundamental Physics (CfFP)
at CCLRC Rutherford Appleton Laboratory.

\newpage
\noindent {\bf Appendix A}

We are making use of a long-established 
result from matrix calculus \cite{horn}, 
whereby, in terms of the matrix derivative $\partial_X=\partial/\partial X$,
for some constant matrix $A$, we have:
\begin{equation}
\partial_X \; {\rm Tr} \; XA= {\rm Tr} \; (\partial_X X) A = A^T \label{theo1}
\end{equation}
where the superscript $T$ denotes the matrix transpose.
Note that $\partial_X X$ is a matrix of matrices
(each sub-matrix with a single unit entry), 
and that in particular $\partial_X X \neq I$,
where $I$ is the identity matrix
(while $\partial_X \, {\rm Tr} \; X = {\rm Tr} \; \partial_X \, X =I$).

We first consider ${\rm Tr} \; C^3 = \, {\rm Tr} \; i[L,N]^3$ 
(where $C:=-i[L,N]$, see Sections~1-2):
\begin{eqnarray}
\partial_L \; {\rm Tr} \, C^3 & = & \; \; {\rm Tr} \; 
            \hspace{4mm} \partial_L \, C^3 \label{a0c31} \\
& = & 3 \; {\rm Tr}
             \hspace{4mm}  (\partial_L C) \; C^2 \label{a0c32} \\
& = & -3i \, {\rm Tr}
             \hspace{4mm}  [(\partial_L L),N] \; C^2 \label{a0c33} \\
& = & -3i \, {\rm Tr}
             \hspace{4mm}  (\partial_L L)[N,C^2] \label{a0c34} \\
& = & -3i \, [N,C^2]^T \label{a0c35}
\end{eqnarray}
making use of the cyclic property of the trace, 
Eq.~\ref{a0c31}-\ref{a0c32} and Eq.~\ref{a0c33}-\ref{a0c34},
and the matrix-calculus theorem (Eq.~\ref{theo1})
for the final step Eq.~\ref{a0c34}-\ref{a0c35}.

The case of ${\rm Tr } \; C^2 = -{\rm Tr } \; [L,N]^2$
(see Section~3) follows entirely analogously:
\begin{eqnarray}
\partial_L \; {\rm Tr} \, C^2 & = & -2i \, [N,C]^T \label{a0c21} \\
& = & -2 \, [N,[L,N]]^T \label{a0c22}
\end{eqnarray}
where in the last line we have
removed the factor $i$ using $C=-i[L,N]$. 

Finally, we consider the case corresponding to $Q_{21} = {\rm Tr} \; [L,N][L^2,N]/2$ (Section~4): 
\begin{equation}
\partial_L \; {\rm Tr} \, [L,N][L^2,N] =  \; {\rm Tr} \, 
            \left( \rule{0mm}{4mm} \, [(\partial_L L),N][L^2,N] 
                                +[L,N]([(\partial_L L)L,N]+[L(\partial_L L),N]) \right) 
                                                    \label{a0211}
\end{equation}
Taking each of the above three terms one at a time, we have:
\begin{eqnarray}
{\rm Tr} \; 
            \hspace{2mm}  \, [(\partial_L L),N][L^2,N]
& = &  \; {\rm Tr}
             \hspace{4mm}  (\partial_L L) [N, [L^2,N]] \label{a0212} \\
{\rm Tr} \; 
            \hspace{2mm}  \, [L,N][(\partial_L L)L,N]
& = &  \; {\rm Tr}
             \hspace{4mm}  (\partial_L L) (LNLN-L^2N^2-LN^2L+LNLN) \hspace{1.0cm} \nonumber \\
& = &  \; {\rm Tr}
             \hspace{4mm}  (\partial_L L) (L[N,[L,N]]) \label{a0213} \\
{\rm Tr} \; 
            \hspace{2mm}  \, [L,N][L(\partial_L L),N]
& = &  \; {\rm Tr}
             \hspace{4mm}  (\partial_L L) (NLNL-LN^2L-N^2L^2+NLNL) \hspace{1.0cm} \nonumber \\
& = &  \; {\rm Tr}
             \hspace{4mm}  (\partial_L L) ([N, [L,N]]L) \label{a0214}
\end{eqnarray} 
wherby adding the above equations:
\begin{eqnarray}
\partial_L \; {\rm Tr} \, [L,N][L^2,N] & = &  \; \; {\rm Tr} \; 
            \hspace{1mm}  \, (\partial_L L) ([N,[L^2,N]]+\{L,[N,[L,N]]\}) \hspace{17mm} \nonumber \\
                  & = & \hspace{2mm} [N,[L^2,N]]^T+\{L,[N,[L,N]]\}^T  \label{a021s}
\end{eqnarray}
where the curly brackets denote the anti-commutator (see main text Eq.~\ref{delq21}).

\newpage
\noindent {\bf Appendix B}

The following completes the expressions for the Lagrange multiplers of Section~4:
\begin{eqnarray}
\lambda_{L00} & = & L_{P2}(12L_3^2L_1 - 6L_3L_2L_1^2 - 18L_3L_2^2 
 \nonumber \\ &   & + 4L_3L_1^4/3 +19L_2^3L_1 - 11L_2^2L_1^3 + 3L_2L_1^5-L_1^7/3) \\
\lambda_{L10} & = & L_{P2}(-36L_3^2 + 36L_3L_2L1 - 4L_3L_1^3 - 3L_2^3 + 3L_2^2L_1^2 -5L_2L_1^4+L_1^6) \\
\lambda_{L20} & = & 4L_{P2}^2(9L_3 - 9L_2L_1 + 2L_1^3) \\
\lambda_{L01} & = &  18L_3^3L_1 - \! 18L_3^2L_2^2 + \! 27L_3^2L_2L_1^2 - \!
                  13L_3^2L_1^4 - \! 48L_3L_2^3L_1 - \! 2L_3L_2^2L_1^3 + \! 14L_3L_2L_1^5
\nonumber \\  &   &  - 2L_3L_1^7 -
       6L_2^5 + 165L_2^4L_1^2/2 - 77L_2^3L_1^4 + 30L_2^2L_1^6 - 6L_2L_1^8 + L_1^{10}/2\\
\lambda_{L11} & = & - 54L_3^3 - 72L_3^2L_2L_1 +
      42L_3^2L_1^3 - 18L_3L_2^3 + 144L_3L_2^2L_1^2 - 74L_3L_2L_1^4
\nonumber \\ &   &  + 22L_3L_1^6/3 -
      6L_2^4L_1 + 10L_2^3L_1^3 - 20L_2^2L_1^5 + 10L_2L_1^7 -4L_1^9/3 \\
\lambda_{L21} & = &  81L_3^2L_2 - 27L_3^2L_1^2 + 108L_3L_2^2L_1 - 90L_3L_2L_1^3 + 18L_3L_1^5
\nonumber \\  &   &   + 27L_2^4/2 - 207L_2^3L_1^2 + 186L_2^2L_1^4 - 56L_2L_1^6 +11L_1^8/2     \\
\lambda_{L02} & = & 36L_3^3L_2 + 36L_3^3L_1^2 - 156L_3^2L_2^2L_1 + 52L_3^2L_2L_1^3  - 16L_3^2L_1^5+ 3L_3L_2^4 
\nonumber \\  &   &  + 93L_3L_2^3L_1^2 - 253L_3L_2^2L_1^4/3 + 35L_3L_2L_1^6 - 4L_3L_1^8 - 41L_2^5L_1/2 
\nonumber \\  &   & + 135L_2^4L_1^3/2 - 62L_2^3L_1^5 + 76L_2^2L_1^7/3 - 11L_2L_1^9/2 + L_1^{11}/2\\
\lambda_{L12} & = & - 144L_3^3L_1 + \! 48L_3^2L_2L_1^2 + \! 32L_3^2L_1^4 - \! 30L_3L_2^3L_1 
                                                    + \! 126L_3L_2^2L_1^3 - \! 254L_3L_2L_1^5/3 
\nonumber \\  &   & + 10L_3L_1^7 + \! 9L_2^5/2 - \! 43L_2^4L_1^2/2 + \! 27L_2^3L_1^4 - \! 25L_2^2L_1^6
                                                             + \! 67L_2L_1^8/6 - \! 3L_1^{10}/2  \\
\lambda_{L22} & = &  216L_3^2L_2L_1 - 72L_3^2L_1^3 - 27L_3L_2^3 - 45L_3L_2^2L_1^2 - 9L_3L_2L_1^4 
\nonumber \\  &   &  + 9L_3L_1^6 + 63L_2^4L_1 - 225L_2^3L_1^3 + 191L_2^2L_1^5 - 59L_2L_1^7 + 6L_1^9\\
\lambda_{L03} & = &  36L_3^4 - 78L_3^3L_2L_1 + 58L_3^3L_1^3 + 24L_3^2L_2^3 -
      63L_3^2L_2^2L_1^2 - 20L_3^2L_2L_1^4/3 - 5L_3^2L_1^6
\nonumber \\  &   &   -
      47L_3L_2^4L_1 + 466L_3L_2^3L_1^3/3 - 298L_3L_2^2L_1^5/3 + 100L_3L_2L_1^7/3 - 11L_3L_1^9/3 
\nonumber \\  &   &  + L_2^6 + 3L_2^5L_1^2/2 -
      15L_2^4L_1^4/2 - L_2^3L_1^6/3 + 7L_2^2L_1^8/3 - 7L_2L_1^{10}/6 + L_1^{12}/6 \\
\lambda_{L13} & = &  18L_3^3L_2 - 150L_3^3L_1^2 + 122L_3^2L_2L_1^3 + 22L_3^2L_1^5/3+ 15L_3L_2^4 
\nonumber \\  &   & - 86L_3L_2^3L_1^2 + 92L_3L_2^2L_1^4 -188L_3L_2L_1^6/3 + 25L_3L_1^8/3
\nonumber \\  &   & - 3L_2^5L_1 + 14L_2^4L_1^3 - 8L_2^3L_1^5 - 2L_2^2L_1^7 + 11L_2L_1^9/3 -2L_1^{11}/3 \\
\lambda_{L23} & = & - 81L_3^2L_2^2 + 270L_3^2L_2L_1^2 -81L_3^2L_1^4 + 108L_3L_2^3L_1 
\nonumber \\  &   &  - 282L_3L_2^2L_1^3 + 104L_3L_2L_1^5 - 22L_3L_1^7/3 - 9L_2^5/2 + 33L_2^4L_1^2/2 
\nonumber \\  &   & - 53L_2^3L_1^4 + 64L_2^2L_1^6 - 49L_2L_1^8/2 +17L_1^{10}/6 \\
\lambda_{L04} & = &  48L_3^4L_1 + \! 24L_3^3L_2^2 - \! 174L_3^3L_2L_1^2 
                                          + \! 166L_3^3L_1^4/3 - \! 91L_3^2L_2^3L_1/2 + \! 1141L_3^2L_2^2L_1^3/6
\nonumber \\  &   &  - 189L_3^2L_2L_1^5/2 + 17L_3^2L_1^7/2 + 5L3L_2^5/4 + 27L_3L_2^4L_1^2/2- 337L_3L_2^3L_1^4/6  
\nonumber \\  &   &  + 89L_3L_2^2L_1^6/3 - 13L_3L_2L_1^8/12 - L_3L_1^{10}/2 + L_2^6L_1/8 - 17L_2^5L_1^3/4 
\nonumber \\  &   &  + 101L_2^4L_1^5/8 - 28L_2^3L_1^7/3 + 61L_2^2L_1^9/24 - 5L_2L_1^{11}/12 +L_1^{13}/24 
\end{eqnarray} 
\begin{eqnarray}
\lambda_{L14} & = & 60L_3^3L_2L_1 - 84L_3^3L_1^3 + 33L_3^2L_2^3/2 - 249L_3^2L_2^2L_1^2/2 + 793L_3^2L_2L_1^4/6 
\nonumber \\  &   & - 19L_3^2L_1^6/2 - 17L_3L_2^4L_1/2 + 139L_3L_2^3L_1^3/3 - 82L_3L_2^2L_1^5/3 
\nonumber \\  &   & - 31L_3L_2L_1^7/3 + 5L_3L_1^9/2 - L_2^6/8 + 9L_2^5L_1^2/4 -67L_2^4L_1^4/8 
\nonumber \\  &   & + 14L_2^3L_1^6/3 - L_2^2L_1^8/24 + 5L_2L_1^{10}/12 -L_1^{12}/8 \\
\lambda_{L24} & = & - 54L_3^3L_2 + 18L_3^3L_1^2 + 54L_3^2L_2^2L_1 + 78L_3^2L_2L_1^3 - 32L_3^2L_1^5 
\nonumber \\  &   & - 27L3L_2^4/4 + 27L_3L_2^3L_1^2 - 293L_3L_2^2L_1^4/2 +199L_3L_2L_1^6/3 - 27L_3L_1^8/4 
\nonumber \\  &   & + 3L_2^5L_1/4 - 5L_2^4L_1^3/2 + 33L_2^3L_1^5/2 - 2L_2^2L_1^7 - 31L_2L_1^9/12 +L_1^{11}/2 \\
\lambda_{L05} & = &  6L_3^4L_2 + 12L_3^4L_1^2 - 19L_3^3L_2^2L_1/2 
                                    - 145L_3^3L_2L_1^3/3 + 29L_3^3L_1^5/2 + L_3^2L_2^4/2 
\nonumber \\  &   & - 5L_3^2L_2^3L_1^2/4 + 2467L_3^2L_2^2L_1^4/36 - 409L_3^2L_2L_1^6/12 + 131L_3^2L_1^8/36 
\nonumber \\  &   &  - L_3L_2^5L_1/6 + 4L_3L_2^4L_1^3 - 223L_3L_2^3L_1^5/6 + 211L_3L_2^2L_1^7/9 -13L_3L_2L_1^9/3 
\nonumber \\  &   & + 2L_3L_1^{11}/9 + L_2^6L_1^2/8 - 11L_2^5L_1^4/6 + 69L_2^4L_1^6/8 
\nonumber \\  &   & - 79L_2^3L_1^8/12 + 137L_2^2L_1^{10}/72 - L_2L_1^{12}/4 +L_1^{14}/72 \\
\lambda_{L15} & = & L_{\Sigma}(-36L_3^3L_1 - 27L_3^2L_2^2/2 + 57L_3^2L_2L_1^2 - 11L_3^2L_1^4/2 + 6L_3L_2^3L_1 
\nonumber \\  &   &  - 15L_3L_2^2L_1^3 - 8L_3L_2L_1^5/3 + L_3L_1^7 -L_2^4L_1^2/2 + 3L_2^2L_1^6/2 - L_2L_1^8/3) \\
\lambda_{L25} & = &  - 18L_3^4 + 36L_3^3L_2L_1 - 4L_3^3L_1^3 
                                              - 3L_3^2L_2^3/4 -141L_3^2L_2^2L_1^2/4
\nonumber \\  &   &   + 111L_3^2L_2L_1^4/4 - 239L_3^2L_1^6/36 -3L_3L_2^4L_1 + 193L_3L_2^3L_1^3/6 
\nonumber \\  &   &   - 87L_3L_2^2L_1^5/2+33L_3L_2L_1^7/2 - 31L_3L_1^9/18 - L_2^6/8 + 2L_2^5L_1^2 
\nonumber \\  &   &   - 81L_2^4L_1^4/8+ 163L_2^3L_1^6/12 - 45L_2^2L_1^8/8 + 3L_2L_1^{10}/4-L_1^{12}/72
\end{eqnarray}
where the invariant traces $L_1$, $L_2$ and $L_3$ 
are defined as in the main text Eq.~\ref{mscons}.
Some useful supplementary polynomials 
(used here and in the main text Eqs.~\ref{rqlm0l}-\ref{rqlm2n}) are:
\begin{eqnarray}
L_{\Sigma} & := & (L_1^3 - L_3)/3 =(m_e+m_{\mu})(m_{\mu}+m_{\tau})(m_{\tau}+m_e) \\
L_{P2} & := & (3L_2 - L_1^2)/2 = m_e^2+m_{\mu}^2+m_{\tau}^2-m_em_{\mu}-m_{\mu}m_{\tau}-m_{\tau}m_e \\ 
L_{P3} & := & (3L_3 - L_2L_1)/2 =m_e^3+m_{\mu}^3+m_{\tau}^3 \nonumber \\        
  & &     \hspace{-3mm} -(m_e^2m_{\mu}+m_{\mu}^2m_e)/2
                 -(m_{\mu}^2m_{\tau}+m_{\tau}^2m_{\mu})/2-(m_{\tau}^2m_e+m_e^2m_{\tau})/2 \\
L_{P6} & := & L_{P3}^2 - L_{\Delta}^2/4 \nonumber
       = (m_e^3+m_{\mu}^3+m_{\tau}^3-m_e^2m_{\mu}-m_{\mu}^2m_{\tau}-m_{\tau}^2m_e) \nonumber \\
      &     & \hspace{2.5cm} \times (m_e^3+m_{\mu}^3+m_{\tau}^3-m_em_{\mu}^2-m_{\mu}m_{\tau}^2-m_{\tau}m_e^2)
\end{eqnarray}
where the charged-lepton discriminant $L_{\Delta}^2$ is defined in the main text, Eq.~\ref{discl}.

\end{document}